\documentclass[prb,aps,twocolumn,floatfix,showpacs]{revtex4}
\usepackage{graphicx}
\usepackage{amsmath}
\usepackage[usenames]{color}

\begin{document}

\title{Correlation effects in the density of states of annealed Ga$_{1-x}$Mn$_{x}$As}

\author{S. Russo}
\affiliation{Kavli Institute of Nanoscience, Delft University of
Technology, The Netherlands}

\author{T.M. Klapwijk}
\affiliation{Kavli Institute of Nanoscience, Delft University of
Technology, The Netherlands}

\author{W. Schoch}
\affiliation{Dept. of Semiconductor Physics, University of Ulm,
Germany}

\author{W. Limmer}
\affiliation{Dept. of Semiconductor Physics, University of Ulm,
Germany}

\date{\today}

\begin{abstract}
We report on an experimental study of low temperature tunnelling
in hybrid NbTiN/GaMnAs structures. The conductance measurements
display a $\sqrt{V}$ dependence, consistent with the opening of a
correlation gap ($\Delta_{\mathrm{C}}$) in the density of states
of Ga$_{1-x}$Mn$_{x}$As. Our experiment shows that low temperature
annealing is a direct empirical tool that modifies the correlation
gap and thus the electron-electron interaction. Consistent with
previous results on boron-doped silicon we find, as a function of
voltage, a transition across the phase boundary delimiting the
direct and exchange correlation regime.
\end{abstract}

\maketitle

\newpage

\begin{figure}[h]
    \centering
    \includegraphics[width=0.8\columnwidth]{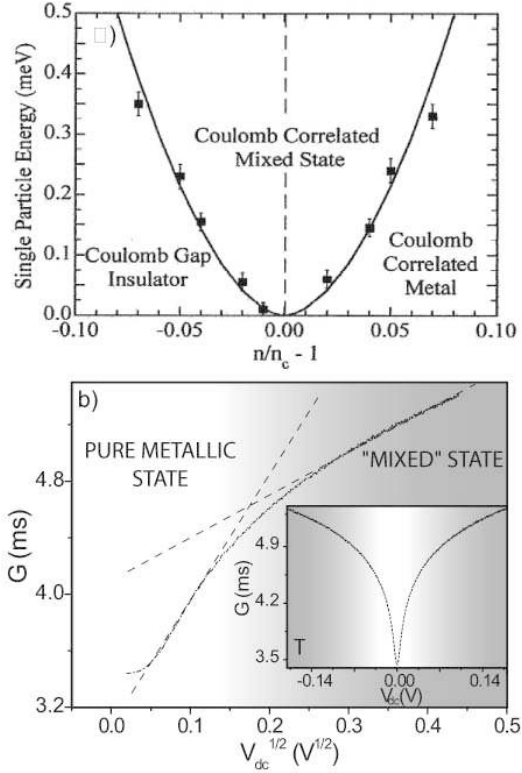}
    \caption {a) Phase-diagram proposed by Lee\cite{Lee} to
    indicate the electron-correlated regimes at low energies and high energies
    as a function of carrier density.
    b) In the inset conductance \textit{versus} voltage bias
    for our NbTiN/GaMnAs devices. Gray gradient in the background
    highlights the transition across the phase boundary delimiting
direct and exchange correlation regime. At low bias
    the superconducting gap of the contact material causes a deviation.}
        \label{Russo_fig1}
\end{figure}

\begin{figure}[h]
    \centering
    \includegraphics[width=0.8\columnwidth]{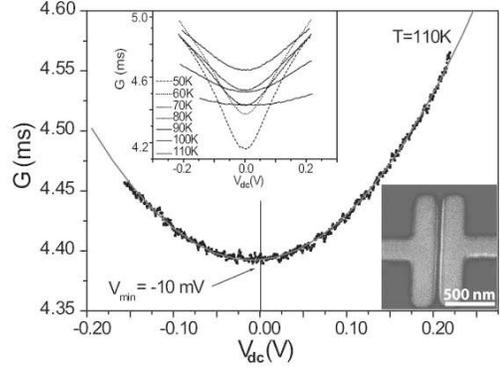}
    \caption  {Central inset: measurements at different temperatures
    of G \textit{versus} V after annealing a device for
    120 min and having a $\mathrm{T}_{\mathrm{Curie}}$ of $96 \mathrm{K}$. Lower inset: a micrograph of one of the
    nano-fabricated S/F/S samples. The main figure shows a conductance measurement
    at T=110K. The full line is a best fit to the BDR model\cite{Brinkman},
    used to determine the barrier height.}
        \label{Russo_fig2}
\end{figure}

\begin{figure}[h]
    \centering
    \includegraphics[width=0.8\columnwidth]{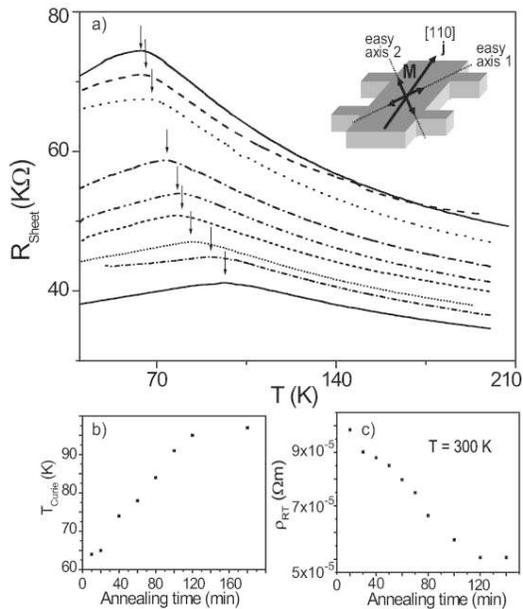}
    \caption  {a) Plot of the sheet resistance versus temperature
    for a GaMnAs Hall bar device, e.g. inset. Different curves are for different annealing
    times, from top to bottom: 0, 10, 20, 40, 60, 80, 100, 120, and 180 min.
T$_{\mathrm{Curie}}$ is highlighted by the arrows. b) The
T$_{\mathrm{Curie}}$ increases linearly with annealing time up to
120 min. and it remains unchanged for further annealing to 180min.
c) The room temperature resistivity decreases as function of
annealing time.}
        \label{Russo_fig3}
\end{figure}

\begin{figure}[h]
    \centering
    \includegraphics[width=0.8\columnwidth]{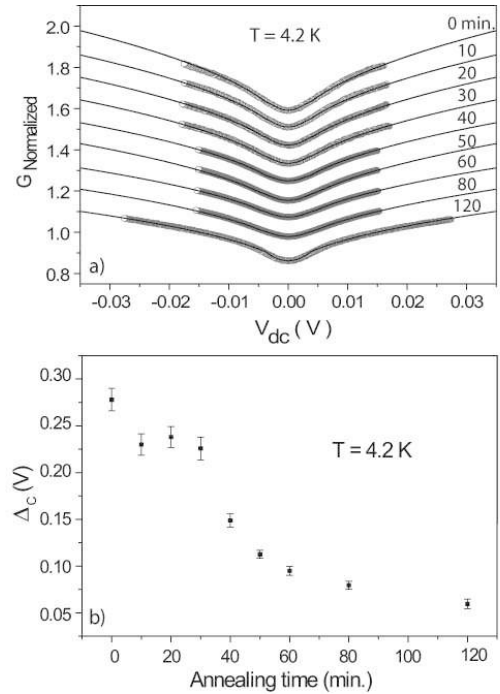}
    \caption  {a) 2-Probe tunnel conductance measurements for different
    low-temperature annealing times (shown for each graph). For increasing
    annealing time the non linear character of the curves at high bias is reduced.
    The solid lines are fits, leading to the correlation gap
    $\Delta_{\mathrm{C}}$.
    b) $\Delta_{\mathrm{C}}$ decreases monotonously with annealing time.}
        \label{Russo_fig4}
\end{figure}

The new class of ferromagnetic semiconductors
Ga$_{\mathrm{1-x}}$Mn$_{\mathrm{x}}$As is known\cite{Matsukura} to
display a metal-insulator transition (MIT) as function of Mn
doping. In conventional doped semiconductors the MIT, which occurs
as a function of carrier density, is widely studied and considered
to be a prime example of a quantum phase transitions. It is
understood that the spatial localization of charge carriers, which
drives the MIT, reduces the ability of the system to screen
charges, leading to a prominent role of the electron-electron
interactions. The experimental trace of the Coulomb interactions
between the electrons is the depletion of the single-particle
Density of States (DOS) $N(E)$ at the Fermi
energy\cite{Shklovskii,Altshuler,McMillan,Massey,Teizer,Chan,Bielejec,Lee}.
For a dirty three dimensional system it is found that $N(E) \ \sim
\ \sqrt{E}$ in the metallic regime\cite{Altshuler,McMillan},
whereas $N(E) \ \sim \ E^2$ in the insulating
regime\cite{Shklovskii}, recently observed in different localized
systems\cite{Massey,Chan,Bielejec}, including
magnetically doped materials\cite{Teizer}.\\
Recently, using conductance measurements across the
metal-insulator-transition Lee\cite{Lee} constructed the phase
diagram shown in Fig. \ref{Russo_fig1}a. At low enough
temperatures, 10mK, the energy is controlled by the voltage at
which the differential conductance is measured. For low energies,
\textit{i.e.} very close to the Fermi energy where the theory for
the MIT is valid the system is a Coulomb gap insulator below the
critical density and a correlated metal above the critical
density. For higher energies a mixed state develops around the
critical density, in which the density of states on both sides of
the transition have a common functional dependence on energies
masking the existence of a critical density. The $\lq\lq$pure"
state at low densities is the regime where exchange correlations
describe the Coulomb interactions, whereas above the critical
density the direct Coulomb interactions rule.  At low energies the
DOS is clearly distinct for metallic and insulating samples and
the system is in the $\lq\lq$pure" state. At high energies the
insulating and metallic states are indistinguishable from
DOS-measurements.\\
The new material system GaMnAs is for low Mn doping an insulator
and the resistivity diverges for $T \rightarrow 0$, indicating
localization effects. In the metallic regime this (III,V)Mn is
characterized by a decreasing resistivity which eventually
saturates for $T \rightarrow 0$, although these resistivity values
remain relatively high ($\sim10^{-3} \ \mathrm{\Omega}$cm, see
Fig. \ref{Russo_fig4}(c)). Thus GaMnAs is a dirty metal where
disorder plays a rather strong role. These strong
electron-electron interaction effects the DOS of GaMnAs\cite{Chun}
and might lead to the observation of the phase boundary cross-over
from direct to exchange correlation at
much higher temperatures than for Si:B.\\
Here we report the observation of the correlation gap in GaMnAs as
measured with a tunnel contact between GaMnAs and the
superconductor NbTiN. At the interface we have a Schottky barrier,
which at low temperatures acts as a tunnel contact, allowing a
direct measurement of the density of states. Superconducting leads
are chosen, on top of unpatterned GaMnAs (inset Fig.
\ref{Russo_fig2}), to ensure that the tunnel junction resistances
are at least an order of magnitude larger than the sample
resistance. The T-shape is chosen to minimize the effect of
parallel conductance paths. With these samples we study
systematically the evolution of the correlation energy
($\Delta_{\mathrm{C}}$) on the annealing time. We find
experimentally that $\Delta_{\mathrm{C}}$ decreases monotonously
with annealing. This behavior suggests that with annealing the
surface of the GaMnAs is driven away from the metallic towards the
insulating state. Furthermore, measurements at bias voltage higher
than $\Delta_{C}$ lead to the observation of a cross-over from the
\textit{direct} correlation regime to the \textit{mixed} state
behavior (as in Fig. \ref{Russo_fig1}b), consistent with previous
results obtained on
Si:B\cite{Lee}.\\
The Ga$_{1-x}$Mn$_{x}$As samples (Mn-content of $4.4 \%$) are
grown on (001) semi-insulating GaAs substrates by low temperature
molecular-beam epitaxy (MBE) at $230 ^\circ \mathrm{C}$. The
GaMnAs epilayer (thickness of  40 nm, $\mathrm{T}_{\mathrm{Curie}}
=64 \mathrm{K}$) is patterned to hold two independent devices: a
Hall bar and the T-shaped tunnel-contacts. The Hall bar
($200\times 50~\mathrm{\mu m}^2$, see inset of Fig.
\ref{Russo_fig3}a) allows the characterization of the magnetic
properties of GaMnAs. Electron Beam Lithography (EBL), Ar RF
sputter cleaning and reactive sputtering are used to define the
top NbTiN (thickness = 30 nm, superconducting transition
temperature $\mathrm{T}_{\mathrm{C}}$ = 15 K and superconducting
gap $\Delta_{\mathrm{S}}$ = $2\mathrm{mV}$). The contacts on the
GaMnAs have a separation of 100 nm and a total area of $0.5\times
1~\mathrm{\mu m}^2$ (see Inset of Fig. \ref{Russo_fig2}). The
tunnel-devices are used to
measure the differential resistance.\\
In the standard tunneling model \cite{Wolf}, the tunneling
conductance $G(V,T) \ = \
\partial I / \partial V$ is the product of the density of states
 in the interacting material, $N_{F}(E)$, with the density of states of
the superconductor, $N_{S}(E)$, convoluted with the
Fermi-distributions. In view of the relevant energies we can
ignore the thermal smearing. $N_{S}(E)$ is given by the standard
BCS density of states as usually modified by a broadening
parameter $\Gamma$\cite{Dynes}: $N_S(E)=N(0)
Re[(E-i\Gamma)/(\sqrt{(E-i\Gamma)^2-{\Delta_{\mathrm{S}}}^2})]$.
The GaMnAs is described as dirty 3D metal
system\cite{Altshuler,McMillan}, thus $N_{F}(E) \ = \
N(0)(1+\sqrt{(E)/(\Delta_{\mathrm{C}})}$. $\Delta_C$ is the
correlation gap which represents the strength of the
electron-electron interaction in the ferromagnetic semiconductor.
In this tunneling description there are two free parameters,
$\Gamma$ and $\Delta_{C}$, while the other parameters are known
independently.\\
For temperatures above the T$_{\mathrm{Curie}}$ the conductance
displays a parabolic dependence on bias voltage\cite{Chun} (see
Fig. \ref{Russo_fig2}). At lower temperatures deviations from the
parabolic behavior occur (see inset in Fig. \ref{Russo_fig2})
which reflect, as we will show, the correlation gap. We focus now
on the 2-probe conductance through the SFS device for
$\mathrm{T}>\mathrm{T}_{\mathrm{Curie}}$, Fig. \ref{Russo_fig2}.
It is apparent that G clearly displays a parabolic dependence on
bias voltage, which demonstrates that tunnelling is taking place,
as described by Brinkman, Dynes and Rowell (BDR)\cite{Brinkman} .
The measured conductance for
$\mathrm{T}>\mathrm{T}_{\mathrm{Curie}}$ shows a slightly
asymmetric shape and the occurrence of a minimum at a finite
voltage bias ($\mathrm{V}_{\mathrm{min}} = -10 \mathrm{mV}$).
These two features in the measurements are typical for the tunnel
conductance in metal-insulator-metal junctions with
different barrier heights at the interfaces.\\
In applying the BDR model to our data to estimate the barrier
height at the S/F interface, we assume that the conduction in the
GaMnAs is mainly due to the heavy holes with an effective mass of
$0.462 \mathrm{m}_0$ \cite{Dietl}. In addition we assume a
thickness of the barrier at S/F of $10 \mathrm{\AA}$\cite{Monch}.
From fitting the curve of $G(V)$ to the BDR model, continuous line
in Fig. \ref{Russo_fig2}, we find that the mean barrier height is
$\overline{\varphi} = 0.33 \mathrm{V}$. Furthermore the bias
voltage at which the minimum conductance occurs
($\mathrm{V}_{\mathrm{min}} = -10 \mathrm{mV}$) gives a difference
in barrier heights at the S/F interface of $\Delta \varphi = 16
\mathrm{mV}$. Finally, we emphasize that the measured resistance
of the two tunnel contacts in series is much higher than the
resistance of the GaMnAs in between. These facts lead us to
conclude that it is reasonable to assume that the measured
conductance is a tunnel conductance.\\
Previous experiments have demonstrated that low temperature post
growth annealing offers the possibility to change the
ferromagnetic properties of GaMnAs\cite{Yu}. Our samples are
annealed in the same way, but performed on fully processed
structure with the NbTiN on top of the GaMnAs. This leaves the
interface unexposed to air. We do not observe any change in the
critical current of the NbTiN, which excludes possible degradation
of the superconductor. The annealing is performed at $200^\circ
\mathrm{C}$ on a hot plate in air \cite{Limmer} for a sequence of
annealing times up to 180 min. In Fig. \ref{Russo_fig3} we present
measurements of the sheet resistance \textit{versus} temperature
performed on the Hall bar for different annealing times. The
resistance displays a non monotonous dependence on temperature. It
reaches a maximum at the Curie temperature\cite{Matsukura} and
eventually decreases, for
$\mathrm{T}<\mathrm{T}_{\mathrm{Curie}}$, as expected for metallic
samples. From the graph of $\mathrm{T}_{\mathrm{Curie}}$
\textit{versus} annealing time (Fig. \ref{Russo_fig3}b) it is
apparent that $\mathrm{T}_{\mathrm{Curie}}$ increases\cite{Ku}. It
remains basically unchanged for further annealing to 180min. which
is consistent with more extensive work presented by Stanciu
\textit{et al}.\cite{Stanciu}. This enhancement of
$\mathrm{T}_{\mathrm{Curie}}$ has been usually traced back to a
removal of compensating defects, and thus, to an increase of the
hole concentration\cite{Yu,Edmonds}. In fact channeling Rutherford
backscattering\cite{Yu} and Auger\cite{Edmonds} experiments have
shown that annealing at low temperature causes a migration of Mn
interstitial defects towards the surface of GaMnAs. The fact that
bulk ferromagnetic properties improve with low temperature
annealing is also evident from the room temperature resistivity
($\rho_{\mathrm{RT}}$), see Fig. \ref{Russo_fig3}c.
$\rho_{\mathrm{RT}}$ decreases monotonously with annealing time,
confirming that a reduction of defects and an
increase of charge density takes place in GaMnAs\cite{Limmer}.\\
The 2-probe tunnel conductance measured at 4.2K is shown in Fig.
\ref{Russo_fig4}a. The measurements are normalized to the
conductance value at 0.015V (arbitrarily chosen) and shifted for
clarity. For $V \ < \ \Delta_{\mathrm{NbTiN}}$ the superconducting
state of the leads dominate the data. However for $V \
> \ \Delta_{\mathrm{NbTiN}}$ correlation effects play the major role in
transport and the measured conductance displays a non-linear
character with the expected $\sqrt{V}$ dependence on the bias
voltage. This $\sqrt{V}$ behavior indicates that the GaMnAs acts
as a three dimensional dirty metal, with correlation effects
parametrized with correlation gap $\Delta_{\mathrm{C}}$. We find
experimentally that the non linear character of the conductance
curves is progressively reduced as a function of increased
annealing time. The continuous lines in Fig. \ref{Russo_fig4}a are
the best fit to the tunneling model. Standard non-linear fitting
is used with the parameters $\Delta_{C}$ and $\Gamma$, and
minimization of the $\chi^2$ merit function is carried out
according to the Levenberg-Marquardt method.  Good agreement
between theory and experiments is found, and for each different
annealing time the corresponding value of $\Delta_{\mathrm{C}}$ is
extracted (see Fig. \ref{Russo_fig4}b); the values found for
$\Gamma$ are 1$\pm$0.2mV.\\
We observe that the interaction parameter in the as grown sample
is $\Delta_{\mathrm{C}} \ =$ 278 mV\cite{nota} and it reduces to a
smallest value of $59 \mathrm{mV}$ by annealing the sample for 120
minutes, see Fig. \ref{Russo_fig4}b. As shown in Fig.
\ref{Russo_fig3} annealing leads to an increase in
T$_{\mathrm{Curie}}$ (from 64K to 97K) and a decrease in
$\rho_{\mathrm{RT}}$ by 48 $\%$, which suggests an improvement in
the quality of material. However, the tunneling measurements lead
to the conclusion that the correlation gap becomes smaller
indicative of a system which is driven from the metallic regime to
a more insulating regime. This behavior is consistent with the
fact that with increasing annealing time a larger number of
compensating defects reaches the surface, causing an increase in
resistivity and a reduction of the correlation energy. Thus low
temperature annealing, while improving the ferromagnetic
properties of the bulk material (see Fig. \ref{Russo_fig2}),
drives the surface of GaMnAs from the
metallic towards the insulating state.\\
We now turn to the tunnel conductance measurements at higher bias
voltage, higher than the correlation gap (V$_{\mathrm{bias}} \
> \ \Delta_{\mathrm{C}}$, e.g. inset Fig. \ref{Russo_fig1}b). We focus on a sample
annealed for 120 minutes and $\Delta_{\mathrm{C}} \ = 59
\mathrm{mV}$. From Fig. \ref{Russo_fig1}b it is apparent that $G \
\sim \ \sqrt{V}$ over the entire bias range but with two different
slopes, one at low energy and a less steep one at higher energy.
The cross-over between these two regimes occurs at the bias
corresponding to the correlation gap. Similar results have been
presented by Lee\cite{Lee}, although at much lower energies.
Adopting the interpretation of Ref.\cite{Lee} we conclude that at
low energies the GaMnAs is properly described as a dirty metal
where correlation effects are manifested in a minimum in the DOS
at the Fermi Energy. However, at high energies the $\sqrt{V}$
dependence stems from a mixture of direct and exchange
correlations. At high energies GaMnAs displays a cross-over to the
\textit{mixed} state. The fact that the energy scale of the
correlation gap in GaMnAs is much higher than in Si:B allowed the
observation of this
cross-over at modest temperatures.\\
In conclusion, we have studied correlation effects in the density
of states  of GaMnAs.  Low temperature post-processing annealing
is found to modify the electron-electron correlation in GaMnAs.
Our experiments suggest that annealing acts in opposite ways on
the bulk compared to the surface of GaMnAs: while improving the
ferromagnetic properties of the bulk it drives the surface from
the metallic state towards the insulating state. Hence, we find
that annealing is a good external parameter which can be used to
monitor continuously the evolution of the correlation gap when
approaching the MIT at the surface of GaMnAs. Interestingly the
tunnel conductance measurements display a cross-over from a low
energy regime to an high energy regime allowing to track the phase
boundary separating the pure metallic behavior from the
\textit{mixed} state, as found previously in Si:B by
Lee\cite{Lee}.\\
The authors acknowledge a useful discussion with S. Rogge. This
work was financially supported by NWO/FOM and the Deutsche
Forschungsgemeinschaft, DFG Li 988/4.


\end{document}